\newif\ifaastex\aastexfalse

\newif\ifdraft\draftfalse

\ifaastex
\documentclass[manuscript]{aastex}
\tabletypesize{\footnotesize}
\newcommand{\deluxetablerotate}{\rotate}
\else
\documentclass{emulateapj}
\renewcommand{\epsscale}[1]{}
\newcommand{\deluxetablerotate}{}
\fi

\newcommand{\sci}{Science}
\newcommand{\rvmp}{RvMP}

\ifdraft
\newcommand{\linenumbers}{}
\newcommand{\nolinenumbers}{}

\usepackage[framemethod=tikz]{mdframed}

\usepackage{color,soul}
\definecolor{myyellow}{rgb}{1,1,0.8}
\sethlcolor{myyellow}

\newcommand{\mynoteB}[1]{\hl{#1}}

\newcommand{\mysoulbox}[1]{\mbox{#1}}

\else

\newcommand{\mynoteB}[1]{#1}
\newcommand{\mysoulbox}[1]{#1}

\newcommand{\linenumbers}{}
\newcommand{\nolinenumbers}{}

\fi

\ifaastex
\else
\let\jjdagger=\dagger
\let\jjddagger=\ddagger
\renewcommand{\dagger}{\ensuremath{\jjdagger}}
\renewcommand{\ddagger}{\ensuremath{\jjddagger}}

\makeatletter
\renewenvironment{deluxetable}[1]{
    \begin{deluxetable*}{#1}
}{
    \end{deluxetable*}
}
\makeatother

\fi

\newcommand{\casa}{Cas~A}
\newcommand{\gtwoninetytwo}{G292.0$+$1.8}

\newcommand{\cmthree}{\ensuremath{\mathrm{cm}^{-3}}}
\newcommand{\kms}{\ensuremath{\mathrm{km\,s}^{-1}}}
\newcommand{\msun}{\ensuremath{\mathrm{M}_{\odot}}}

\newcommand{\NH}{\ensuremath{N_{\mathrm{H}}}}

\newcommand{\hmss}[4]{#1$^{\mathrm h}$ #2$^{\mathrm m}$ #3$\fs$#4}

\newcommand{\dmss}[4]{#1$\arcdeg$ #2$\arcmin$ #3$\farcs$#4}

\shorttitle{RSG wind of \casa}
\shortauthors{Lee et al.}

\begin{document}

\title{X-ray observation of the shocked red supergiant wind of Cassiopeia A}

\author{Jae-Joon Lee\altaffilmark{1,2} Sangwook Park\altaffilmark{3},
{John P. Hughes\altaffilmark{4}},
and {Patrick O. Slane\altaffilmark{5}}
}

\altaffiltext{1}{Korea Astronomy and
Space Science Institute, Daejeon 305-348, Korea}
\altaffiltext{2}{leejjoon@kasi.re.kr}
\altaffiltext{3}{Department of Physics, University of Texas at Arlington
Arlington, TX 76019}
\altaffiltext{4}{Department of Physics and Astronomy, Rutgers University, 136
Frelinghuysen Road, Piscataway, NJ 08854-8019}
\altaffiltext{5}{Harvard-Smithsonian Center for Astrophysics, 60 Garden Street,
Cambridge, MA 02138}

\linenumbers

\begin{abstract}
  \casa\ is a Galactic supernova remnant whose supernova explosion is
  observed to be of Type IIb from spectroscopy of its light echo
  \citep{2008Sci...320.1195K}.  Having its SN type known,
  observational constraints on the mass loss history of Cas~A's progenitor
  can provide crucial information on the
  final fate of massive stars.  In this paper, we study X-ray
  characteristics of the shocked ambient gas in \casa\ using the 1 Msec
  observation carried out with the {\em Chandra X-ray Observatory}
  \citep{2004ApJ...615L.117H}, and try to constrain the mass loss
  history of the progenitor star.
  We identify
  thermal emission from the shocked ambient gas along the outer
  boundary of the remnant.
  Comparison of
  measured radial variations of spectroscopic parameters of the
  shocked ambient gas to the self-similar solutions of
  \citet{1982ApJ...258..790C} show that Cas~A is expanding into a
  circumstellar wind rather than into a uniform medium.
  We estimate a wind density $n_{\mathrm{H}} \sim 0.9\, \pm 0.3\,\cmthree$ at
  the current outer radius of the remnant ($\sim 3$ pc), which we
  interpret as a dense slow wind from a red supergiant (RSG) star.
  Our results suggest that the progenitor star of \casa\ had an initial
  mass around $16$ \msun, and its mass before the explosion was about
  $5$ \msun, with uncertainties of several tens of
  percent. Furthermore, the results suggest that, among the
  mass lost from the progenitor star ($\sim$11 \msun), a significant
  amount (more than 6 \msun) could have been via
  its RSG wind.
\end{abstract}

\keywords {ISM: individual (\casa) --- supernova remnants ---
  X-rays: ISM ---stars: winds}

\section{Introduction}

The mass of a star is the most crucial parameter that governs
its structure, evolution and fate. Stellar masses vary
throughout the life of stars: stars are subject to mass loss by
winds and may experience mass exchange with their companion stars.
For massive stars, in particular, mass
loss plays a critical role in their evolution
\citep[e.g.,][]{Maeder00_361.p159}, but this is a poorly understood part of
 stellar evolution theory.  Mass loss also profoundly impacts
the eventual supernova (SN) explosion \citep{RevModPhys.74.1015}
and the consequent evolution of the supernova remnant (SNR).
Massive stars, with their strong ionizing radiation and powerful winds,
clear out a large area around them throughout their lifetime
\citep[e.g., a radius larger than 15 pc for stars earlier than
B0;][]{1999ApJ...511..798C}.
This implies that the immediate environment of a massive star
is thus determined by the mass loss history of the star itself. It
further implies that when the massive
star
undergoes core-collapse and explodes as a supernova, the supernova
shocks expand inside the circumstellar structure.
In other words, the structure and the
evolution of ``young'' core-collapse SNRs reflects the
pre-supernova mass-loss history of the progenitors, providing
observational constraints on the nature of their progenitor stars.

Observations have established that Type~II-P SNe are produced when
stars with initial masses in the range 8 -- 17\ \msun\ undergo core
collapse during their red supergiant (RSG) phases
\citep{Smartt09_MNRAS395.p1409}.
\mynoteB{
On the other hand,
the origin of other types of SNe is still not well established.
Single-star evolutionary models such as those by
{\mysoulbox{\citet{2003ApJ...591..288H}}}
predict stars with initial masses $\gtrsim
30\,\msun$ will completely shed their hydrogen envelopes to become
Wolf-Rayet (WR) stars, and explode as Type~Ib/c SNe. The observed
statistics of SNe types, however, does not seem consistent with such
models {\mysoulbox{\citep[e.g., see the discussion in][]{Smith11_MNRAS412.p1522}}}.
Of particular interest is the fate of stars with initial masses of 17 --
30 {\msun}, as they do not seem to produce Type~II-P SNe.
Some of them may explode during the RSG phase
producing Type~II-L, IIb, or IIn SNe.  But observational constraints
are indirect and weak.  Also, the fate of
massive stars may be complicated by binary interactions.  }

For a star that undergoes a SN explosion during the RSG phase (or soon
after it evolved past the RSG phase), its remnant will primarily
interact with the surrounding RSG wind.  Existence of such an RSG wind
has been suggested for SN~1987A
\citep{1989ApJ...342L..75C,2005ApJ...627..888S},
\mynoteB{
where its RSG phase was followed by a blue supergiant phase during which
the SN explosion occurred.
}
The extent of the
RSG wind would be determined by the location where the ram pressure of
the wind equals the pressure of the surrounding medium. A radius of 5
pc is estimated for a canonical case \citep[][$\dot{M}=5\times 10^{-5}
\,\msun\,\mathrm{yr}^{-1}$, $v_w = 15\,\kms$, $p/k=10^{4} \cmthree$
K]{2005ApJ...619..839C}.  Thus, the interaction of the SNR with the
RSG wind can last for thousands of years.
The Galactic
core-collapse SNR \gtwoninetytwo\ is likely an example that
is currently
interacting with the RSG wind of its progenitor
\citep[][LEE2010
hereafter]{2010ApJ...711..861L}. By analyzing the X-ray emission from
the outer shock region of this $\sim3000$ year-old SNR with the
\emph{Chandra} X-ray observatory, we found that the remnant has been
expanding inside a wind.  The estimated wind density ($n_{\mathrm{H}}
= 0.1 \sim 0.3\ \cmthree$) at the current outer radius ($\sim7.7$ pc)
of the remnant is consistent with a slow wind from an RSG star.

\casa, one of the best studied Galactic SNRs, also appears to be
interacting with its RSG wind. At its age of about 330 yrs
\citep[][and references therein]{2001AJ....122..297T}, its morphology
and expansion rates are consistent with a model SNR interacting with a
RSG wind \citep{2003ApJ...593L..23C}.  It is also found that the
observed characteristics of the X-ray ejecta knots are consistent
with \casa\ expanding into a wind
\citep{2003ApJ...597..347L,2009ApJ...703..883H}.  On the other hand,
the direct identification of shocked RSG wind has been limited.  The
remnant contains slow moving shocked circumstellar clumps, called
quasi-stationary flocculi (QSF). While earlier studies suggested that
these could be fragments of the RSG shell swept-up by the later
Wolf-Rayet wind, more recent studies favor that these are more likely
dense clumps in the RSG wind \citep{2003ApJ...593L..23C}.  While
observations have revealed detailed physical properties of these QSFs,
the global characteristics of the RSG wind have not been well
established so far.

The characteristics of the wind surrounding \casa, i.e., the mass loss
history of the progenitor of \casa, is particularly pertinent as we
know the supernova spectrum of \casa\ at the time of its explosion
from the light echo observations
\citep{2008Sci...320.1195K,Rest11_732.p3}.  The SN spectrum of \casa\
shows that it is remarkably similar to that of SN~1993J, a prototypical
Type~IIb SN whose progenitor is likely an RSG star
\citep{1993Natur.364..507N}.

In this paper, we study the circumstellar environment of \casa\ and
try to constrain the mass loss history of the progenitor star.  The
basic method is similar to the one we applied to \gtwoninetytwo\ in LEE2010,
although here we use the 1 Msec \emph{Chandra} observation of \casa\
\citep{2004ApJ...615L.117H}.  The data are briefly summarized in
Section~\ref{sec:data}.  In Section~\ref{sec:analysis}, we describe
our detection of thermal X-ray emission from the shocked ambient gas
and study its spectral characteristics.  In
Section~\ref{sec:discuss-nature}, we show that the observed
characteristics of the shocked ambient gas are consistent with those
expected from a shocked RSG wind.  Then the implication of our
results on the nature of the progenitor star
(Section~\,\ref{sec:progenitor-casa}) and cosmic ray acceleration
(Section~\ref{sec:discuss-cr}) is discussed.
Section~\ref{sec:summary-nature} summarizes our results.

\section{Data}
\label{sec:data}

\casa\ was the first light target of the \emph{Chandra} X-ray
Observatory \citep{2000ApJ...528L.109H},
and has been observed multiple times with \emph{Chandra}.
A one million second exposure of \casa\ was obtained with the ACIS-S3
detector on board \emph{Chandra} in 2004 \citep{2004ApJ...615L.117H}.
We use this one million second \emph{Chandra} observation in our study. The 1 Msec
observation data consist of 9 separate ObsIDs (4634, 4635, 4636, 4637,
4638, 4639, 5196, 5319 an 5320) with their exposure time ranging
between $\sim$40 ks and $\sim$170 ks.  We reprocessed the level 1
event files to create new level 2 event files. We applied parameters
of the standard \emph{Chandra} pipeline process.
CIAO 4.5 and CALDB 4.5.7 were used for all
the reprocessing and analysis.

\section{Analysis}
\label{sec:analysis}

\subsection{Imaging Analysis}
\label{sec:imaging-anaylsis}

\begin{figure*}[htb]
  \centering
  \plotone{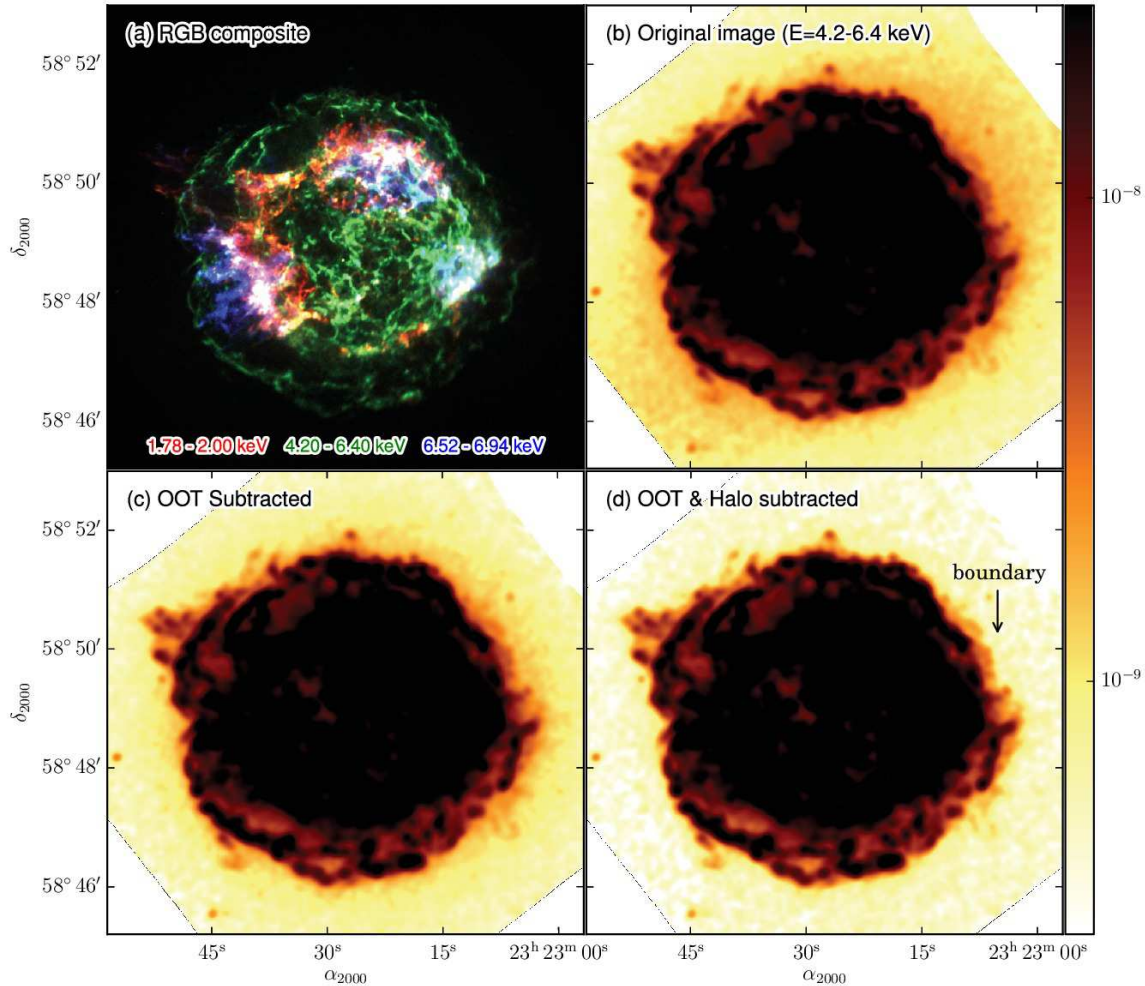}
  \caption{ \emph{(a)} Three-color image of \casa\ with
    red for 1.78$-$2.0 keV, blue for 6.52$-$6.95 keV, and green for
    4.2$-$6.4 keV. The energy range
    for each color is adopted from Figure 1 of \citet{2004ApJ...615L.117H}.
    \emph{(b)} Unprocessed \emph{Chandra} image (4.2--6.4 keV) of
    \casa. The image is smoothed with Gaussian beam of FWHM=4.5\arcsec
    for better visualization.  \emph{(c)} Same \emph{Chandra} image as in the
    left but with the out-of-time events subtracted (see the text for
    details). \emph{(d)} Same \emph{Chandra} image as in the center but the
    estimated dust scattered halo emission subtracted (see the text
    for details). This final image reveals the existence of faint
    emission (orange--ish color outside saturated black)
    that extends beyond the non-thermal filaments, especially
    in northwestern region. A boundary of this faint emission in the
    east is indicated with an arrow. \mynoteB{The color scale is selected to
    better reveal the faint structures,
    and bright features in the inner area are mostly saturated to black.}
  }
  \label{fig:oosub-step}
\end{figure*}

The X-ray emission from \casa\ is dominated by emission from shocked
metal rich ejecta (Figure~\ref{fig:oosub-step}(a)).
Nonthermal filaments are seen along the outer boundary of \casa,
which are synchrotron emission from the TeV
electrons accelerated at the shock front \citep[][for a recent
review]{2008ARA&A..46...89R}.  On the other hand, our study focuses on
diffuse thermal emission from the shocked circumstellar gas that is
expected to be fainter than the nonthermal emission.
Although non-thermal X-ray  filaments are seen
along most of the outer boundaries, there are regions with weak
non-thermal emission, especially in the northwest. The X-ray emission
in these regions shows no prominent limb-brightening, and we suspect
that this is emission from shocked circumstellar wind.

\begin{figure}[hb]
  \centering
  \plotone{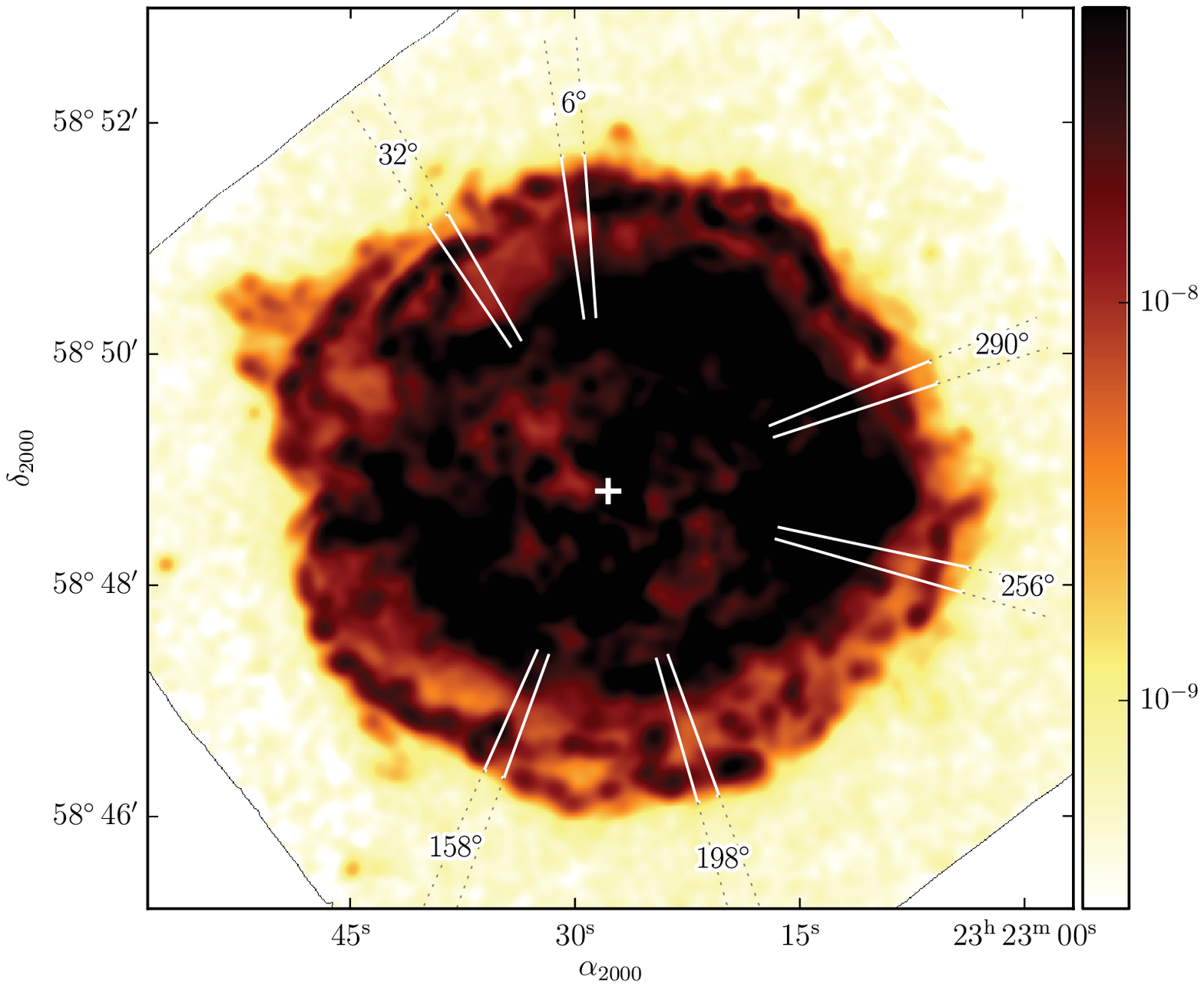}
  \caption{ Background subtracted \emph{Chandra} image (4.2--6.4 keV) of \casa\
    smoothed with Gaussian beam of FWHM=4.5\arcsec. This is the same image
    as Figure~\ref{fig:oosub-step}(d),
    but with a slightly different color scale.
    Solid white lines
    are the areas where the radial profiles in
    Figure~\ref{fig:radial-profiles} are extracted.
    They are shown as thin black dashed lines outside
    the remnant for clarity. Associated labels mark their
    position angles. The plus sign marks the adopted explosion
    center of Cas A \citep[$\alpha_{\mathrm{2000}}$ = \hmss{23}{23}{27}{77},
$\delta_{\mathrm{2000}}$ = \dmss{58}{48}{49}{4};][]{2001AJ....122..297T}.
   }
  \label{fig:oosub-final}
\end{figure}

The identification of the X-ray emission from shocked wind in \casa\
is complicated not only due to non-thermal filaments but also due to
contamination from ejecta emission. Furthermore, for the faint
emission from shocked circumstellar gas, non--uniform backgrounds could be
problematic. We consider two background components that are
spatially varying, thus hindering the identification of the
emission from the shocked circumstellar gas; they are out-of-time events
\footnote{\url{http://cxc.harvard.edu/proposer/POG/html/chap6.html\#sec:trailed-images}}
along the CCD readout directions and a dust-scattered X-ray halo
around bright sources \citep{1965ApJ...141..864O}. In \casa, both
background components originate primarily from the bright ring of the
shocked metal-rich ejecta, whose radius is about three quarters of the
outer radius of the remnant.  In the following, we describe our
attempt to remove these two non-uniform background components. We
note, however, that our intention of removing background components is
to improve the ``imaging'' analysis of the faint emission from
the outer shock region. For our spectral analysis, we use
original data.

To remove the contamination from out-of-time (OOT) events, we create
OOT event files from level 1 event files following the method outlined
in the M.~Markevitch's script
make\_readout\_bg\footnote{\url{http://cxc.harvard.edu/contrib/maxim}}.
In short, this method substitutes chip-y (the readout direction)
values of each event in the original level 1 event list with random
ones, and processes them identically to the original event files (we
refer them as OOT event files).  From these OOT event files, we create
OOT images.  Then, the OOT images are subtracted from the image
created using the original event files scaled for the ratio of the
readout time over the frame time (0.041/3.2), resulting in the
OOT-subtracted images.

The second background component considered is the dust-scattered X-ray
halo emission, which is formed around an X-ray source due to small
angle scattering by foreground dust particles
\citep{1989ApJ...336..843M,1995A&A...293..889P}.
Physical modeling of such emission, especially for an extended source,
is complicated.  It depends on the distribution of dust along the
line-of-sight and the energy of the X-ray photons.  Instead,
a simple approach is adopted, where the spatial distribution of the
halo emission
is modeled as Gaussian profile around the source.
In other words, the
dust-scattered halo emission is estimated by ordinary
Gaussian-convolution of the source image.
\mynoteB{
The OOT-subtracted image is
smoothed with a Gaussian kernel of a certain width, then scaled and
subtracted from the original OOT-subtracted image.  The width of
the kernel and the scaling factor were adjusted until the resulting
image has the most flat background. The flatness is simply measured
as a standard deviation of regions well outside the remnant.
}
This
approach gives a poor approximation of the real halo
emission near the source. However, we find that this provides a
reasonable approximation for the outer regions that we are interested
in.

The steps of background subtraction are illustrated in
Figure~\ref{fig:oosub-step}.
The dominant background contribution in the original image
(Figure~\ref{fig:oosub-step}(b)) is the OOT contribution seen along
the direction of northwest to southeast which is the readout
direction. The OOT contribution is effectively removed in the
OOT-subtracted image (Figure~\ref{fig:oosub-step}(c)) while the
remaining halo emission is clearly visible.
And we see that the halo contribution is significantly reduced in our
final image (Figure~\ref{fig:oosub-step}(d)).
\mynoteB{
This final image clearly reveals the existence of faint emission
that extends beyond the non-thermal filaments, especially in
northwestern region.
We note that the results depend on
the chosen energy band to some degree and better subtraction was
obtained when high energy bands above 3 keV are used.  However, as long
as the energy range is high enough to be insensitive to foreground
absorption and broad enough to be dominated by continuum emission,
the results are similar regardless of the chosen energy range.
In this paper,
the energy range of 4.2 -- 6.4 keV is adopted.
}

\begin{figure}[t]
  \centering
  \epsscale{0.5}
  \plotone{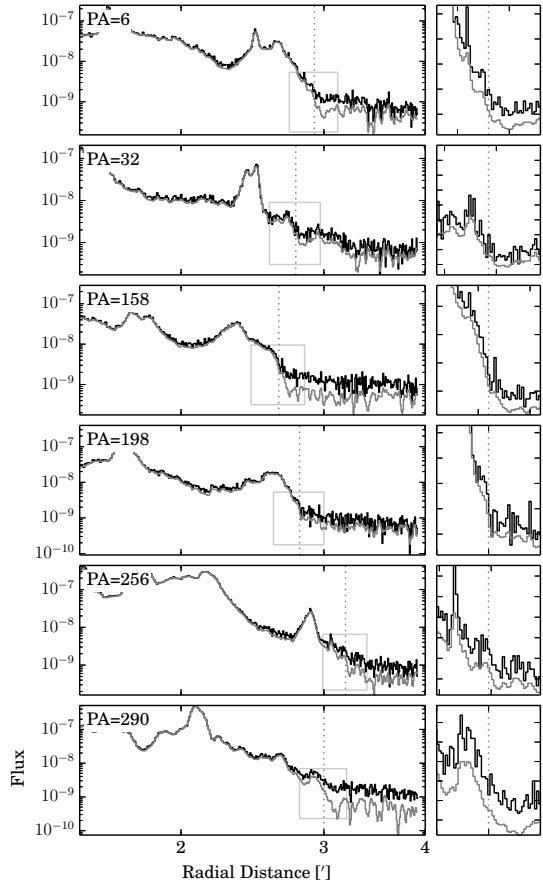}
  \caption{ Radial intensity profiles (E=$4.2 - 6.4$ keV)
    extracted along the several regions shown in
    Figure~\ref{fig:oosub-final}. The black lines are from the original
    image and the gray lines are from the background subtracted image. The
    left panels show the radial profile in a log-log scale. The right
    panels show the same radial profiles in a linear-linear scale
    for the region near the outer boundary (marked as gray boxes on the left
    panes). The locations of the outer
    boundary visually identified from Figure~\ref{fig:oosub-final} are
    marked as vertical dashed lines.}
  \label{fig:radial-profiles}
  \epsscale{1.}
\end{figure}

\mynoteB{
The faint emission and its boundary is not likely an artifact of
our background subtraction as our background subtraction only
affects large scale structures. For example, the width of the
Gaussian kernel applied to remove dust--scattered halo is about
1\arcmin.
To further test if the faint emission is real, we extract radial
intensity profiles in several regions from the
original image and also from the background-subtracted image.
}
In Figure~\ref{fig:oosub-final}, we reproduce 
Figure~\ref{fig:oosub-step}(d) in false-color with different contrast.
We also mark the explosion center
($\alpha_{\mathrm{2000}}$ = \hmss{23}{23}{27}{77},
$\delta_{\mathrm{2000}}$ = \dmss{58}{48}{49}{4}) of the remnant
from \citet{2001AJ....122..297T}, which is adopted as the
remnant center throughout this paper.
White lines in Figure~\ref{fig:oosub-final} are the regions where the
radial profiles are extracted
(these regions are associated with the regions where spectra
are extracted in Section~\ref{sec:spectral-anaylsis}), and their radial
profiles
are shown in
Figure~\ref{fig:radial-profiles}.
For each profile in Figure~\ref{fig:radial-profiles}, we mark the
locations of the outer shock front that we visually identified from
the background-subtracted image (Figure~\ref{fig:oosub-final}).  We
find that the profiles extracted from the original image do show
changes in their slope across the proposed boundaries, indicating they
are real.  As will be discussed in the next section, spectra of these
regions confirm their ambient origin.

\mynoteB{
Figure~{\ref{fig:oosub-final}} (and Figure~{\ref{fig:oosub-step}}) reveal
the existence of faint
diffuse emission marking the outer boundary of \casa.
We note that this emission, in projection, often
extends beyond the nonthermal filaments (bright filamentary features
near the boundary).
The nonthermal emission is relatively weak along the western boundary
where the emission shows a faint plateau.  In contrast to the
non-thermal filaments, little hint of limb-brightening is seen in this
region. In fact, the surface brightness tend to increase inwardly from
the boundary, if the contribution of nonthermal filaments is
neglected, as demonstrated in Figure~{\ref{fig:radial-profiles}}.
For example, the radial profile for position angle of 256{\arcdeg}
clearly indicates its surface brightness gradually increase from its suggested
boundary around 3\farcm2 inwardly to radius of 2\arcmin.  The local
peak around 2\farcm9 is due to the nonthermal filament.
A similar
behavior can be noticed from other radial profiles in
Figure~{\ref{fig:radial-profiles}}.
Thus, the emission shows
no significant limb-brightening and its surface brightness
increases inward from
the outer boundary. These characteristics are
expected from the thermal
emission of the shocked ambient gas when the remnant is expanding into
a wind, as demonstrated in the case of {\gtwoninetytwo}
(LEE2010).
Therefore, this faint structure provides promising
evidence of emission features from a shocked circumstellar wind.
}

\subsection{Spectral Analysis}
\label{sec:spectral-anaylsis}

\begin{figure}[h]
  \centering
  \plotone{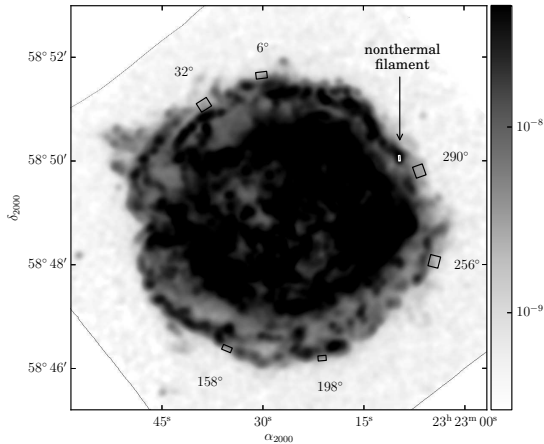}
  \caption{ Regions from which the azimuthal spectra in
    Figure~\ref{fig:azimuthal-spectra} are extracted. The regions are
    labeled by their position angles from the remnant center.}
  \label{fig:azimuthal-spectra-region}
\end{figure}

\begin{figure}[h]
  \centering
  \epsscale{0.7}
  \plotone{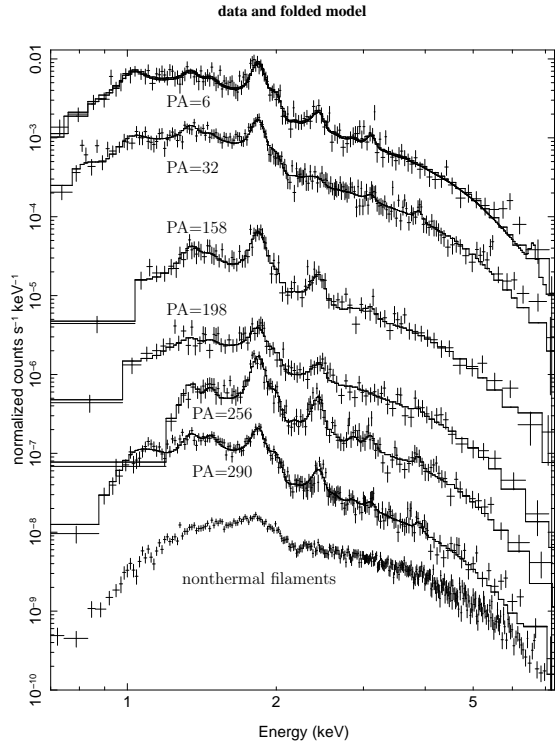}
  \caption{ Extracted \emph{Chandra} spectra of the outer faint regions in
    Figure~\ref{fig:azimuthal-spectra-region} and
    their best fit models.  The bottom spectrum is from the nonthermal
    filament also marked in Figure~\ref{fig:azimuthal-spectra-region},
    shown for comparison.
}
  \label{fig:azimuthal-spectra}
  \epsscale{1.}
\end{figure}

We extract spectra from several regions
around the outermost boundary
believed to be
dominated by thermal emission of shocked ambient gas.
The extraction regions are shown in
Figure~\ref{fig:azimuthal-spectra-region} and the corresponding
spectra are shown in
Figure~\ref{fig:azimuthal-spectra}.
We selected regions with little contamination from the
ejecta emission and the nonthermal emission.
\mynoteB{
The spectra are extracted from the original event files, thus the processes
described in Section~\,{\ref{sec:imaging-anaylsis}} do not
apply for the spectral analysis.
Instead, the background spectrum is extracted from a
nearby region outside the remnant boundary, typically about 10\arcsec away
from the source region.
The scale length of the halo emission near the boundary, estimated
from the width of the Gaussian kernel we used to
subtract the scattering halo, is about 1\arcmin.
Therefore, local background spectra (within 10\arcsec) will be
effective in removing the halo component from the source spectra.
The OOT contribution is significant only if the readout direction
of the source region intersect with bright emission in the remnant
center. The OOT contribution for spectra for PA=6, 32, 198, 256 should
be insigficant. For spectra for PA=290 and 158, their background regions
are selected along their readout direction so that they have
a similar OOT contribution as the corresponding source regions.
Therefore, contributions from the OOT event and the halo
component should be insignificant in our spectral analysis.
}
The spectra of these regions clearly show silicon and sulfur lines.
The overall spectral slope in the hard parts of our spectra is also softer than
that of the nonthermal filament (an example spectrum of
nonthermal filament is also shown in
Figure~\ref{fig:azimuthal-spectra} for a comparison).  Therefore, we
conclude that the X-ray emission we have extracted is dominated by a thermal
component.

The extracted spectra are fit with a single nonequilibrium ionization
plane-parallel shock model \citep[vpshock in Xspec
v12,][]{2001ApJ...548..820B} with varying abundances. The spectra are
not binned; instead we use Churazov
weighting \citep{1996ApJ...471..673C}.
The net photon count for each spectrum is
between 5,000 and 20,000 photons.
The best fit models are overlaid in
Figure~\ref{fig:azimuthal-spectra}.
We find that a single vpshock model adequately fits the observed
spectra.  The fitted abundances are similar to or less than the
solar values.
A similar low abundance pattern has been also found from the outer
shock of \gtwoninetytwo\ (LEE2010). Therefore we suggest that the faint emission
we see is thermal emission from shocked ambient gas.
In Table~\ref{tab:az-fit}, we summarize the fit parameters for each
regions, sorted by their position angle (P.A.).
It is possible that there is some contribution from
the nonthermal emission,
and any residual
nonthermal component would affect the fit parameters.
We note that fitting the spectra with additional powerlaw component
(with its photon index fixed at 2.5 from the fit to nearby nonthermal filaments)
decreases the fitted temperature while increasing the ionization time scale,
but other parameters remain consistent within their error ranges.

\newcommand{\TableAzCaption}{X-ray spectral parameters of the regions
  used in the azimuthal analysis. \label{tab:az-fit}}
\newcommand{\PaNote}{Position angles to each
  region are measured east of north from the geometrical center.}
\newcommand{\DofNote}{For all individual spectra, the degree of
  freedom ($\nu$) is 454.}
\newcommand{\TableAzComments}{\DofNote}
\begin{deluxetable}{ccccccccccccc}
\deluxetablerotate
\tablecolumns{13}
\tablewidth{0pc}
\tablecaption{\TableAzCaption}
\tablehead{
\colhead{P.A.\tablenotemark{\dagger}} & \colhead{Area} & \colhead{$N_{\mathrm{H}}$} & \colhead{$kT_e$} & \colhead{norm} & \colhead{$n_e t$} & \colhead{Ne} & \colhead{Mg} & \colhead{Si} & \colhead{S} & \colhead{Ca} & \colhead{Fe} & \colhead{reduced $\chi^2$\tablenotemark{\ddagger}}\\
\colhead{[$^{\circ}$]} & \colhead{$\arcsec^2$} & \colhead{[$10^{22}$ cm$^{-2}$]} & \colhead{[keV]} & \colhead{[10$^{-19}$ cm$^{-5}$]} & \colhead{[$10^{11}$ cm$^{-3}$s]} & \colhead{} & \colhead{} & \colhead{} & \colhead{} & \colhead{} & \colhead{} & \colhead{}
}
\startdata
{6} & {399} & {0.9$_{-0.1}^{+0.1}$} & {2.3$_{-0.3}^{+0.3}$} & {9.7$_{-1.5}^{+2.1}$} & {1.9$_{-0.5}^{+0.7}$} & {0.2$_{-0.2}^{+0.3}$} & {0.2$_{-0.1}^{+0.1}$} & {0.4$_{-0.1}^{+0.1}$} & {0.3$_{-0.1}^{+0.1}$} & {0.0$_{-0.0}^{+0.6}$} & {0.2$_{-0.1}^{+0.2}$} & {0.90}\\
{32} & {655} & {1.0$_{-0.1}^{+0.1}$} & {2.1$_{-0.2}^{+0.1}$} & {22.7$_{-1.8}^{+1.5}$} & {1.5$_{-0.3}^{+0.3}$} & {0.0$_{-0.0}^{+0.1}$} & {0.1$_{-0.0}^{+0.1}$} & {0.3$_{-0.1}^{+0.1}$} & {0.1$_{-0.1}^{+0.1}$} & {0.8$_{-0.7}^{+0.7}$} & {0.1$_{-0.1}^{+0.0}$} & {1.03}\\
{158} & {269} & {1.9$_{-0.3}^{+0.3}$} & {2.2$_{-0.4}^{+0.4}$} & {8.6$_{-1.9}^{+2.0}$} & {1.5$_{-0.3}^{+0.5}$} & {0.0$_{-0.0}^{+1.7}$} & {0.5$_{-0.2}^{+0.5}$} & {0.7$_{-0.2}^{+0.2}$} & {0.5$_{-0.2}^{+0.2}$} & {0.0$_{-0.0}^{+1.2}$} & {0.9$_{-0.5}^{+0.5}$} & {1.01}\\
{198} & {228} & {1.2$_{-0.2}^{+0.5}$} & {2.3$_{-0.4}^{+0.4}$} & {8.1$_{-1.6}^{+3.2}$} & {1.9$_{-0.8}^{+1.9}$} & {0.2$_{-0.2}^{+0.5}$} & {0.1$_{-0.1}^{+0.1}$} & {0.2$_{-0.1}^{+0.1}$} & {0.2$_{-0.2}^{+0.8}$} & {0.3$_{-0.3}^{+0.8}$} & {0.0$_{-0.0}^{+0.4}$} & {0.94}\\
{256} & {645} & {1.6$_{-0.2}^{+0.3}$} & {2.2$_{-0.3}^{+0.4}$} & {17.3$_{-3.0}^{+5.6}$} & {2.0$_{-0.4}^{+0.6}$} & {0.0$_{-0.0}^{+0.8}$} & {0.5$_{-0.2}^{+0.2}$} & {0.9$_{-0.2}^{+0.3}$} & {0.7$_{-0.2}^{+0.1}$} & {0.9$_{-0.9}^{+0.8}$} & {0.0$_{-0.0}^{+0.2}$} & {1.11}\\
{290} & {655} & {1.3$_{-0.1}^{+0.2}$} & {1.8$_{-0.2}^{+0.1}$} & {34.7$_{-3.9}^{+5.8}$} & {4.0$_{-0.9}^{+0.9}$} & {0.0$_{-0.0}^{+0.5}$} & {0.4$_{-0.1}^{+0.2}$} & {0.5$_{-0.1}^{+0.1}$} & {0.3$_{-0.1}^{+0.1}$} & {0.6$_{-0.5}^{+0.5}$} & {0.4$_{-0.1}^{+0.2}$} & {1.02}\\
\enddata
\tablenotetext{\dagger}{\PaNote}
\tablenotetext{\ddagger}{\DofNote}
\end{deluxetable}

\begin{figure}
  \centering
  \epsscale{0.7}
  \plotone{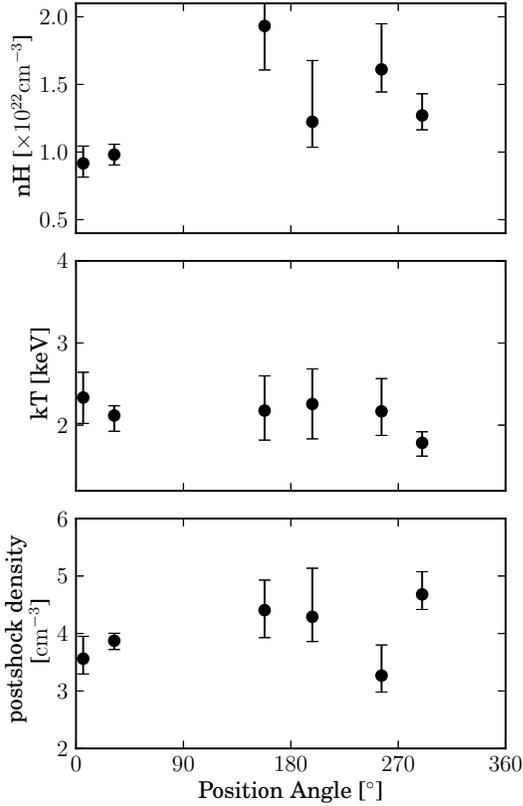}
  \caption{ Azimuthal distribution of the absorbing column densities
    (top), electron temperatures (middle) and the postshock densities
    (bottom) from the spectral fits of the outermost regions in
    Figure~\ref{fig:azimuthal-spectra}.}
  \label{fig:azimuthal-variation}
  \epsscale{1.}
\end{figure}

\begin{figure}
  \centering
  \epsscale{0.7}
  \plotone{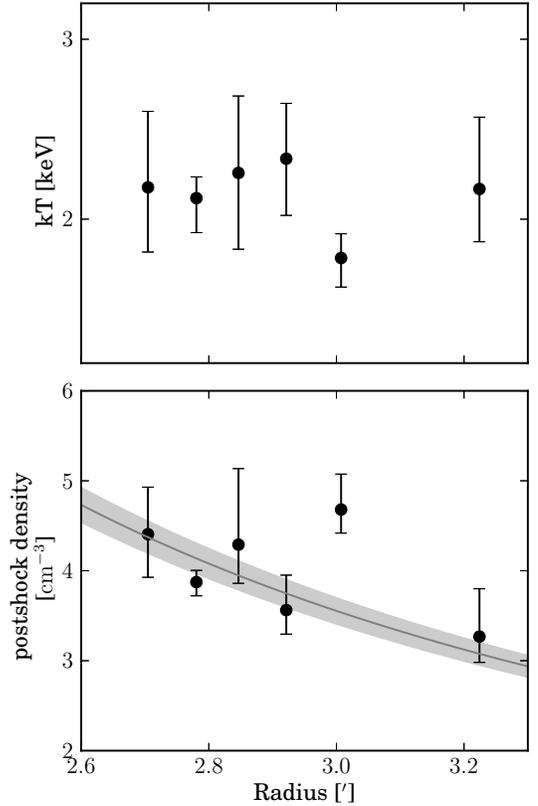}
  \caption{ Electron
    temperatures (top) and the postshock densities (bottom)
    from the spectral fits of the outermost regions in
    Figure~\ref{fig:azimuthal-spectra}
    are plotted as a function of their radial distances from
    the center. If \casa\ is expanding inside a wind, the radial variation of
    postshock density in this figure will follow the preshock
    density profile scaled by the shock compression ratio.
    The solid line is the best fit model with a simple wind profile
    ($\rho \propto r^{-2}$) and the gray area represents the formal
    error of the fit for 90\% confidence interval.}
  \label{fig:azimuthal-radial-variation}
  \epsscale{1.}
\end{figure}

In Figure~\ref{fig:azimuthal-variation}, we compare spectral
parameters of each region as a function of their position angles.
The absorbing hydrogen column densities and electron temperatures are
given from the spectral fit. The densities are estimated from the
fitted volume emission measures. The volume corresponding to each
region is calculated by pixel-by-pixel integration of the path-lengths
over the projected area. In this calculation, we assumed a sphere
centered at the assumed center of the remnant at a distance of 3.4 kpc
\citep{1995ApJ...440..706R}. The radius of the sphere is separately
determined for each region as the projected distance of the farthest
(from the assumed center of the remnant) pixel in the region.
The absorbing hydrogen column densities seem relatively low in the
north compared with other directions, which seems consistent with
the results of \citet{2012ApJ...746..130H}.
On the other hand, we could not see a clear azimuthal
trend in the electron temperatures and the hydrogen
densities. Similarly, Figure~\ref{fig:azimuthal-radial-variation}
shows variation of the electron temperatures and the hydrogen
densities of each region as a function of their distance from the
remnant center.
No clear
indication of radial trends is noticed although
the radius range that we are sampling is rather limited
($2.7\arcmin \lesssim$ R $\lesssim 3.2\arcmin$).

\begin{figure}[h]
  \centering
  \plotone{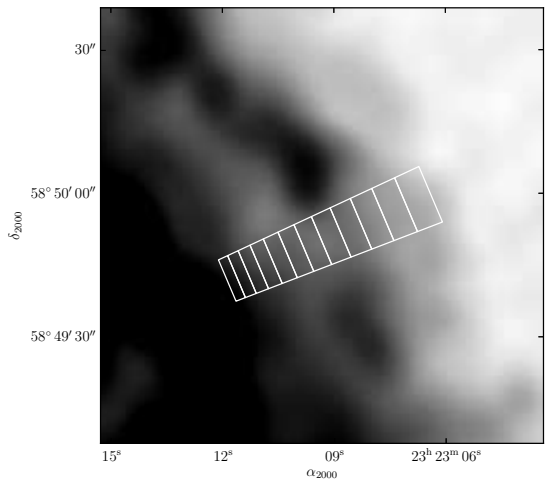}
  \caption{ Zoomed view of the northwestern part of
    Figure~\ref{fig:oosub-final} with regions for the radial spectral
    analysis marked as white boxes.  }
  \label{fig:radial-profiles-regions}
\end{figure}

\begin{figure}[h]
  \centering
  \plotone{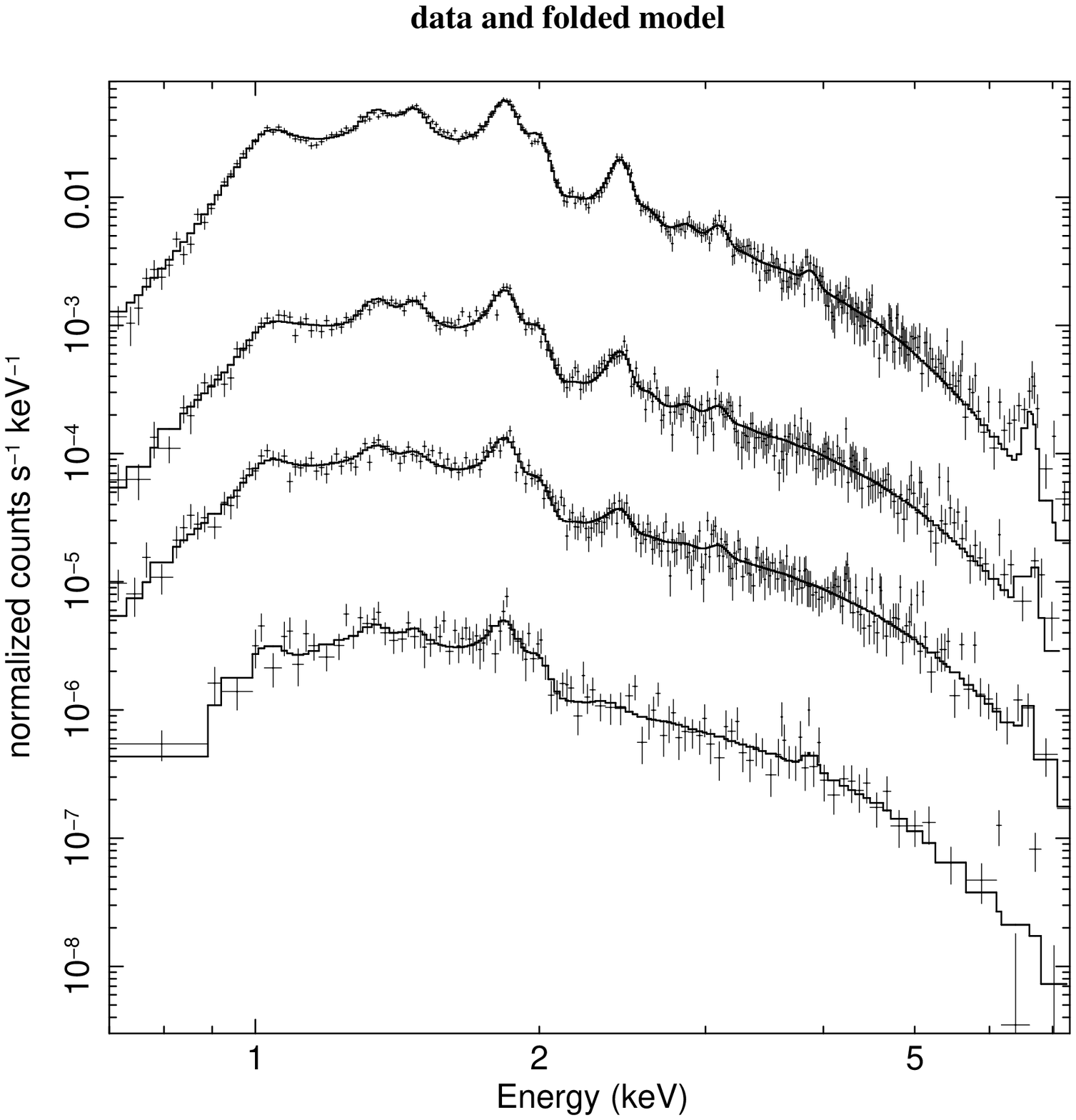}
  \caption{ Sample of \emph{Chandra} spectra extracted from regions for the
    radial analysis in Figure~\ref{fig:radial-profiles-regions}. The
    uppermost spectra corresponds to the innermost region, and the
    lowermost spectra to the outermost region. Their best fit models
    are also shown. }
  \label{fig:radial-spectra}
\end{figure}

\begin{figure}[h]
  \centering
  \epsscale{0.6}
  \plotone{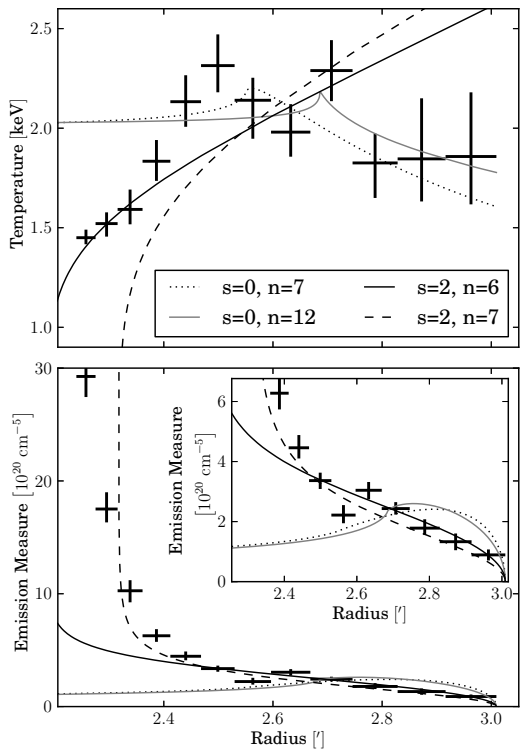}
  \caption{ Fitted electron temperatures and emission measures for the
    radial series of spectra from Figure~\ref{fig:radial-spectra},
    plotted as a function of their radial distances from the center.
    Lines are model predictions
    from
    \citeauthor{1982ApJ...258..790C} (\citeyear{1982ApJ...258..790C};
    see Section~\ref{sec:discuss-nature} for more details) for
    different sets of parameters, where $s$ and $n$ are power-law
    indices for the density profile of the ambient medium and of the
    expanding ejecta, respectively.
    The temperatures and emission measures of the model predictions
    have been scaled to match the observed values.
    The inset of the bottom panel shows the same data with different scales.
    For the models of $s=0$, the contact discontinuity is marked by cusps in
    the temperature profile.
    For the models of $s=2$, the contact discontinuity is where the temperature
    drops to zero and the emission measure diverges to infinity.
}
  \label{fig:radial-model}
  \epsscale{1.0}
\end{figure}

As has been demonstrated in the case of
G292.0+1.8 (LEE2010), one can investigate
the radial structure of the ``shocked'' ambient gas to test whether
the gas is of interstellar origin or of circumstellar origin.  We note
that the radial structure we are considering here is qualitatively
different from what is shown in
Figure~\ref{fig:azimuthal-radial-variation}, which represents
characteristics of ``multiple'' shocks of different directions (and at
different radii from the center). By radial structure, we mean radial
variations of gas characteristics behind ``a'' shock, which represents
accumulated history of shock propagation.
Investigating the nature of ambient gas from the radial structure
behind a shock requires a radial series
of spectra that covers large enough radial range to test different
models.
We select regions in the northwestern boundary where emission from the
nonthermal filaments is insignificant.
Figure~\ref{fig:radial-profiles-regions} shows the regions
where we extract the radial series of spectra.
We allow the area of the regions to increase outwardly to compensate
for decreasing surface brightness.
Again, the background spectra are extracted from the regions
right outside the shock boundary. In the inner region away from the shock
boundary, the contribution of the OOT
events and the halo emission may remain in the source spectra.
However, we do not think this significantly affects our analysis because,
as will be shown later, the emission
from the CSM itself is much stronger
in this inner region.
The spectra of individual regions are first fit
independently as in the azimuthal analysis.  The fitted metal
abundances are found to be less than those of solar values, even in
the bright innermost region well inside the remnant boundary. This
suggests that most emission (if not all) of our selected
regions is from shocked ambient gas. The fitted absorbing
column densities (\NH) of these regions were consistent
within their uncertainties.  We consider that the variation of \NH\ in
this small patch of the sky is negligible compared to uncertainties in the
fit parameters and we refit the whole set of spectra with \NH\ fixed at
the value of $1.2\times 10^{22}\ \mathrm{cm}^{-2}$, an average from
the independent fits.
Figure~\ref{fig:radial-spectra} shows a selected sample of the spectra
from these regions and their best fit models.
And the fit results are summarized in Table~\ref{tab:rad-fit} and
plotted in Figure~\ref{fig:radial-model}.
From Table~\ref{tab:rad-fit}, one can easily notice that
the ionization time scale ($n_e \times t$) increases significantly
with distance inward from the shock.
As such a trend is
expected for material behind a shock that is moving outward,
this confirms that
the emission is from shocked ambient gas.
It is also clear that the observed emission measure monotonically increases
toward the remnant center.
On the other hand, the variation of the electron temperature
is complicated.  In the inner regions (radius smaller than
$2.5\arcmin$), we see a clear tendency of inwardly decreasing electron
temperature. But the temperature remains more or less flat in the
outer regions (radius between $2.5\arcmin \sim 3.0\arcmin$).  For some
elements (e.g., Ne, Mg, and Si), their abundances are higher for the
inner regions. However, it is not clear whether this trend is real or
is due to some systematic uncertainties.
The radial structure behind the shock will be further discussed in the
next section.

\newcommand{\TableRadCaption}{X-ray spectral parameters of the regions
  used in the radial analysis. \label{tab:rad-fit}}
\renewcommand{\DofNote}{For all individual spectra, the degree of
  freedom ($\nu$) is 455.}
\newcommand{\TableRadComments}{}
\begin{deluxetable}{cccccccccccc}
\deluxetablerotate
\tablecolumns{12}
\tablewidth{0pc}
\tablecaption{\TableRadCaption}
\tablehead{
\colhead{Distance} & \colhead{Area} & \colhead{$kT_e$} & \colhead{norm} & \colhead{$n_e t$} & \colhead{Ne} & \colhead{Mg} & \colhead{Si} & \colhead{S} & \colhead{Ca} & \colhead{Fe} & \colhead{reduced $\chi^2$\tablenotemark{\dagger}}\\
\colhead{[$^{'}$]} & \colhead{$\arcsec^2$} & \colhead{[keV]} & \colhead{[10$^{-19}$ cm$^{-5}$]} & \colhead{[$10^{11}$ cm$^{-3}$s]} & \colhead{} & \colhead{} & \colhead{} & \colhead{} & \colhead{} & \colhead{} & \colhead{}
}
\startdata
{2.22} & {69} & {1.5$_{-0.1}^{+0.1}$} & {85.4$_{-3.8}^{+6.6}$} & {9.2$_{-1.2}^{+1.6}$} & {0.7$_{-0.2}^{+0.2}$} & {0.8$_{-0.1}^{+0.1}$} & {0.8$_{-0.1}^{+0.1}$} & {0.7$_{-0.1}^{+0.1}$} & {0.6$_{-0.3}^{+0.3}$} & {0.4$_{-0.1}^{+0.1}$} & {1.24}\\
{2.26} & {82} & {1.4$_{-0.1}^{+0.1}$} & {92.5$_{-5.8}^{+5.0}$} & {10.6$_{-1.4}^{+1.3}$} & {1.0$_{-0.2}^{+0.3}$} & {0.8$_{-0.1}^{+0.1}$} & {0.7$_{-0.1}^{+0.1}$} & {0.7$_{-0.1}^{+0.1}$} & {0.6$_{-0.6}^{+0.8}$} & {0.3$_{-0.1}^{+0.1}$} & {1.29}\\
{2.30} & {97} & {1.5$_{-0.1}^{+0.1}$} & {65.8$_{-4.5}^{+5.6}$} & {6.0$_{-0.7}^{+1.0}$} & {1.0$_{-0.2}^{+0.3}$} & {0.7$_{-0.1}^{+0.1}$} & {0.6$_{-0.1}^{+0.1}$} & {0.6$_{-0.1}^{+0.1}$} & {0.4$_{-0.3}^{+0.3}$} & {0.3$_{-0.1}^{+0.1}$} & {1.36}\\
{2.34} & {111} & {1.6$_{-0.1}^{+0.1}$} & {45.6$_{-4.5}^{+4.2}$} & {5.6$_{-0.8}^{+1.0}$} & {1.1$_{-0.3}^{+0.2}$} & {0.8$_{-0.1}^{+0.2}$} & {0.8$_{-0.1}^{+0.1}$} & {0.6$_{-0.1}^{+0.1}$} & {0.8$_{-0.8}^{+0.8}$} & {0.4$_{-0.1}^{+0.1}$} & {0.99}\\
{2.39} & {125} & {1.8$_{-0.1}^{+0.1}$} & {31.9$_{-2.7}^{+3.0}$} & {4.1$_{-0.6}^{+0.8}$} & {0.7$_{-0.3}^{+0.3}$} & {0.5$_{-0.1}^{+0.1}$} & {0.6$_{-0.1}^{+0.1}$} & {0.5$_{-0.1}^{+0.1}$} & {0.3$_{-0.3}^{+0.8}$} & {0.4$_{-0.1}^{+0.1}$} & {1.01}\\
{2.44} & {142} & {2.1$_{-0.1}^{+0.1}$} & {25.5$_{-2.0}^{+2.4}$} & {3.7$_{-0.6}^{+0.7}$} & {0.4$_{-0.3}^{+0.3}$} & {0.6$_{-0.1}^{+0.1}$} & {0.6$_{-0.1}^{+0.1}$} & {0.5$_{-0.1}^{+0.1}$} & {0.0$_{-0.1}^{+0.4}$} & {0.4$_{-0.1}^{+0.1}$} & {1.01}\\
{2.50} & {160} & {2.3$_{-0.1}^{+0.2}$} & {21.9$_{-1.8}^{+1.7}$} & {3.8$_{-0.6}^{+0.8}$} & {0.5$_{-0.3}^{+0.4}$} & {0.6$_{-0.1}^{+0.1}$} & {0.5$_{-0.1}^{+0.1}$} & {0.3$_{-0.1}^{+0.1}$} & {0.2$_{-0.2}^{+0.5}$} & {0.4$_{-0.1}^{+0.1}$} & {1.19}\\
{2.56} & {177} & {2.1$_{-0.2}^{+0.1}$} & {16.3$_{-2.0}^{+2.4}$} & {3.1$_{-0.6}^{+0.9}$} & {0.7$_{-0.4}^{+0.5}$} & {0.4$_{-0.1}^{+0.2}$} & {0.5$_{-0.1}^{+0.1}$} & {0.2$_{-0.1}^{+0.1}$} & {0.0$_{-0.1}^{+0.5}$} & {0.4$_{-0.1}^{+0.1}$} & {0.94}\\
{2.63} & {197} & {2.0$_{-0.1}^{+0.1}$} & {24.7$_{-2.4}^{+2.2}$} & {2.6$_{-0.4}^{+0.5}$} & {0.0$_{-0.1}^{+0.3}$} & {0.2$_{-0.1}^{+0.1}$} & {0.3$_{-0.1}^{+0.1}$} & {0.2$_{-0.1}^{+0.1}$} & {0.2$_{-0.2}^{+0.5}$} & {0.4$_{-0.1}^{+0.1}$} & {1.01}\\
{2.71} & {217} & {2.3$_{-0.2}^{+0.2}$} & {22.0$_{-1.8}^{+1.9}$} & {2.2$_{-0.4}^{+0.5}$} & {0.2$_{-0.2}^{+0.3}$} & {0.2$_{-0.1}^{+0.1}$} & {0.3$_{-0.1}^{+0.1}$} & {0.2$_{-0.1}^{+0.1}$} & {0.0$_{-0.1}^{+0.3}$} & {0.3$_{-0.1}^{+0.1}$} & {1.01}\\
{2.79} & {240} & {1.8$_{-0.2}^{+0.1}$} & {17.7$_{-2.3}^{+2.9}$} & {3.6$_{-1.1}^{+1.8}$} & {0.9$_{-0.6}^{+0.8}$} & {0.3$_{-0.1}^{+0.2}$} & {0.3$_{-0.1}^{+0.1}$} & {0.1$_{-0.1}^{+0.1}$} & {0.2$_{-0.2}^{+0.7}$} & {0.2$_{-0.1}^{+0.1}$} & {0.96}\\
{2.87} & {260} & {1.8$_{-0.2}^{+0.3}$} & {14.6$_{-3.1}^{+3.0}$} & {2.8$_{-0.8}^{+1.6}$} & {0.8$_{-0.5}^{+0.8}$} & {0.3$_{-0.1}^{+0.2}$} & {0.3$_{-0.1}^{+0.1}$} & {0.3$_{-0.1}^{+0.2}$} & {0.9$_{-0.8}^{+0.8}$} & {0.1$_{-0.1}^{+0.2}$} & {0.91}\\
{2.96} & {288} & {1.9$_{-0.2}^{+0.3}$} & {10.6$_{-2.2}^{+2.2}$} & {3.8$_{-1.7}^{+4.4}$} & {0.7$_{-0.7}^{+1.2}$} & {0.3$_{-0.3}^{+0.8}$} & {0.3$_{-0.3}^{+0.8}$} & {0.0$_{-0.1}^{+0.2}$} & {0.8$_{-0.8}^{+1.0}$} & {0.1$_{-0.1}^{+0.2}$} & {0.96}\\
\enddata
\tablenotetext{\dagger}{\DofNote}
\end{deluxetable}

\section{Nature of the Ambient Gas}
\label{sec:discuss-nature}

The radial structure of the shocked ambient gas can trace the radial
density structure of the medium into which the remnant is expanding.
\citet{1982ApJ...258..790C} found self-similar solutions for the
interaction of expanding stellar ejecta ($\rho \propto r^{-n}$, $n>5$)
with an external medium ($\rho \propto r^{-s}$) where $s$ and $n$ are
the power law indices of the density profile for ambient gas and for the
ejecta, respectively: a remnant expanding in a uniform ambient medium
($s=0$) would result in an inward-decreasing density and an
inward-increasing temperature profile of shocked ambient gas,
while a remnant expanding in a
medium with a wind profile ($s=2$) would show an inward-increasing
density and an inward-decreasing temperature.  The primary effect
of $n$ (the density structure of the ejecta) to the solution is a change in
the distance between the forward shock and the contact discontinuity,
and does not fundamentally change the overall behavior.  Therefore, by
studying the current radial profile of the density and temperature of
shocked ambient gas, one can infer whether the SNR has been expanding
in a density gradient.  The utility of this method has been
successfully demonstrated in our study of Galactic core-collapse SNR
\gtwoninetytwo\ (LEE2010).

\mynoteB{
For \casa, previous studies suggested that the remnant is expanding
within its wind. {\mysoulbox{\citet{2003ApJ...593L..23C}}} finds its morphology
and expansion rates being consistent with a model SNR interacting with
a RSG wind.
This is also supported by the observed characteristics
of the X-ray ejecta knots {\mysoulbox{\citep{2003ApJ...597..347L,2009ApJ...703..883H}}}.
The X-ray surface brightness in
Figure~{\ref{fig:radial-profiles}} show a gradual increase
inward from the outer boundary. The increase of surface brightness
is likely due to the increase in emission measure, which is expected
when the remnant is expanding within a wind ($s=2$).
This can be more directly confirmed from our radial spectral analysis.
}
In Figure~{\ref{fig:radial-model}}, we overlay model predictions for the radial structure
of the temperature and emission measure on top of the
observed ones. We adopt the self-similar solutions of
\citet{1982ApJ...258..790C} and project them assuming a
spherical geometry.
As in LEE2010, the projected temperature
is taken as an average temperature along
the line of sight weighted by density squared, ignoring the variation
in the intrinsic emissivity of the gas at different temperatures.
The increasing emission measure toward the remnant center
is the behavior expected for
the remnant expanding inside a wind and we find that the observation
is well represented by self-similar solutions with $s=2$ and $6 < n < 7$.
The temperature variation is not consistent with
any of the models, although the decrease in temperature in the inner part (R $\lesssim$ 2.5 pc) is what is expected in the $s=2$ models.
As will be discussed below, direct comparison
between the observed and predicted temperatures is difficult due to
complicated shock physics, especially near the shock front. On the
other hand, the fact that the observed variation of the emission
measure is in good agreement with models of $s=2$ clearly suggest that
the density of the ambient gas radially decreases, which confirms that
\casa\ has been expanding inside the wind.  While our data suggest a
power-law index $n$ of ejecta between 6 and 7, this may not be
a reliable estimate as has been discussed in LEE2010. We note,
however, that \citet{2003ApJ...597..347L} studied ejecta emission in
\casa\ and found $n$ between 6 to 7.

The temperatures from our X-ray spectral fits are
electron temperatures, while the temperatures of the models are
ion temperatures. These two can differ considerably in
non-radiative shocks faster than 1000 \kms\ \citep[e.g.,][and
references therein]{2007ApJ...654L..69G}.
The shock velocity measured from the proper motion of nonthermal
filaments is about 5000 \kms\ \citep{2004ApJ...613..343D}, while the
measured electron temperature ranges between 2 to 2.5 keV
(Table~\ref{tab:az-fit}). This
corresponds to $\mathrm{T}_{\mathrm{e}}/\mathrm{T}_{\mathrm{p}}$ = 0.05,
assuming all the shock energy has
converted into the energy of the ordinary gas (i.e., no cosmic ray
acceleration; however, see Section~\ref{sec:discuss-cr}).
Previous studies suggested that, for fast shocks ($v_s
\gtrsim 1000\ \kms$), the temperature equilibration between electrons
and protons are low ($\mathrm{T}_{\mathrm{e}}/\mathrm{T}_{\mathrm{p}}
\lesssim 0.1$) and possibly scales with $v^{-2}$
\citep{2007ApJ...654L..69G}.
The observed
electron temperature in \casa\ and its inferred
$\mathrm{T}_{\mathrm{e}}/\mathrm{T}_{\mathrm{p}}$, assuming no cosmic
ray acceleration, seems to be consistent with observed values of other
SNRs. Since it is known that there is CR acceleration
in \casa, it is likely that the observed low electron
temperature is partially due to an efficient cosmic ray acceleration
at the forward shock \citep[e.g., ][]{2003ApJ...584..758V}.
Nevertheless, the measured electron temperature is likely
not to trace the temperature expected from the self-similar solutions
of \citet{1982ApJ...258..790C} and
the discrepancy between the models and the observed values likely
represent incomplete physics of the self-similar models we used.
However, while efficient cosmic ray acceleration may affect some of
our results, e.g,
estimated preshock densities, we do not consider the effects
to be highly significant in the sense that they will not invalidate
our primary result that \casa\ has been expanding inside its wind.

The properties of shocked ambient gas we estimated in
Section~\ref{sec:spectral-anaylsis} are therefore the properties of
wind material.
The measured postshock densities in
Figure~\ref{fig:azimuthal-radial-variation} are from regions around
the outermost boundary for different position angles,
and the self-similar solutions of \citet{1982ApJ...258..790C} show
that the density remains flat near the shock front. Thus, we conclude that these
measured densities closely approximate wind densities at the location
of the forward shock,
scaled by the assumed postshock compression ratio.
\mynoteB{
We emphasize that the postshock density is not affected by
asymmetry of the ejecta and is only dependent on the wind density (and
compression ratio).
We note however that the wind might have not been
completely spherical
and its
characteristics may vary gradually from pole to equator.
}

Fitting the
observed densities in Fig.~\ref{fig:azimuthal-radial-variation}(b)
with a \mynoteB{simple spherically symmetric} wind profile assuming a
constant postshock compression ratio of 4
gives a
wind density profile of
\[
n_{\mathrm{H}} = 0.89 \left( \frac{r}{3\, \mathrm{pc}} \right)^{-2} \cmthree.
\]
The formal error from the fit (for 90\% confidence interval) is about
5\%. However, given the scatter of density values for individual
regions, which may reflect the asymmetric nature of the wind, an error
of 30\% would be more realistic.
The inferred density profile
of \casa\ corresponds to the mass loss rate of
\[\dot{M} = 2.5 \times
10^{-5}\ (\frac{v_w}{10\,\kms})\, \msun\,\mathrm{yr}^{-1}.\] where
$v_w$ is the wind velocity.  The mass loss rate is too large to be
that of the WR wind ($\dot{M} > 10^{-3}\, \msun\,\mathrm{yr}^{-1}$ for
$v_w > 1000\, \kms$) and consistent with that of the RSG wind
estimated from SNe studies \citep[e.g.,][]{2003LNP...598..145S}.
Therefore, our results support that \casa\ is currently expanding
within the RSG wind.  Using the morphology and expansion rates of
\casa, \citet{2003ApJ...593L..23C} estimated explosion energy ($E$)
and the total mass of the ejecta ($M_{e}$), but as a function of the
wind density.  Substituting our observed wind density, we get $E=5
\times 10^{51}$ erg and $M_{e} = 4\, \msun$.  The mass of the swept-up
wind is calculated by the volume--integration of the wind density
profile within the current radius of the remnant, which is around $6\,
\msun$.  For all these estimates, their uncertainties appear to be
dominated by the uncertainties of the numerical coefficients in the
relation derived by \citet{2003ApJ...593L..23C}, which were suggested
to be $\sim50\%$,  and these estimates may be further subject to
systematic uncertainties.  For example, these values are calculated
assuming the RSG wind had a constant mass loss rate, and will be
affected by any possible time variation of the mass loss rate.  It was
proposed that the dense wind of Cas~A may not have extended directly
to the surface of the progenitor star and that \casa\ might have exploded
into a low density bubble \citep[e.g.,][]{2008Sci...320.1195K}.
However, \citet{2008ApJ...686..399S} and \citet{2009ApJ...703..883H}
found that the size of the bubble, if it existed, would not have been
larger than 0.3 pc, which is significantly smaller than the current
radius of \casa\ ($\sim3$ pc). Therefore, the effect of the bubble on
our estimates does not seem significant.  \citet{1996A&A...307L..41V}
and \citet{1997A&A...324L..49F} derived a total swept up mass ranging
from 5 to 9 \msun\ using the global X-ray spectrum of \casa.  While
this is in agreement with ours, the mass estimates based on the fits
to the global X-ray spectrum may suffer from uncertainties due to the
spatial variation of X-ray spectra and the contribution from shocked
ejecta.  
\citet{2012ApJ...746..130H} independently estimated an explosion energy of
$2.4 \times 10^{51}$ erg and ejecta mass of $3.1\, \msun$, both values of 
which are smaller than our nominal estimates. However, the estimated
uncertainties in our estimates based on the approximate solutions from
\citet{2003ApJ...593L..23C}
are as large as 50\%. Coupled with uncertainties
in the total X-ray emitting mass and filling factor used in the
estimates by \citet{2012ApJ...746..130H}, we consider the two results to be
compatible.

\section{Progenitor of \casa}
\label{sec:progenitor-casa}

A lower limit on the progenitor's initial mass can be estimated by
combining all the masses associated with \casa. If we combine
the shocked wind mass ($\gtrsim 6\ \msun$), the ejecta mass ($4\
\msun$), and the mass of the neutron star ($\sim 1.4\ \msun$), we
estimate the initial mass of the progenitor to be larger than
$11\ \msun$. Another estimate of the initial mass comes from the SN
type of \casa. The SN spectra of \casa\ observed by the light echos
show that they can be classified as arising from a Type~IIb SN
\citep{2008Sci...320.1195K,Rest11_732.p3}. Such SNe originate from the
core collapse of massive stars that have lost most of their hydrogen
envelopes and consist of a nearly bare helium core at the time of its
collapse. Thus the He core mass of the progenitor is supposed to be
comparable to the stellar mass at the time of explosion,
which is estimated as $\sim5\ \msun$
by summing the masses of the ejecta and the neutron star. 
The He core mass of $\sim5\ \msun$ implies an initial
mass of $\sim16$ \msun\ \citep[e.g.,][]{RevModPhys.74.1015} with
possibly several tens of percent of uncertainties.  We note that this
mass is close to the upper limit of Type~II-P SNe
\citep[$\sim15 \msun$,][]{Smartt09_MNRAS395.p1409}.
For a comparison, the progenitor
star of the SN~1993J, the prototypical SN Type~IIb, was a red
supergiant \citep[][and references therein]{1993Natur.364..507N}
with a main sequence mass
probably in between 13 and 20 \msun \citep[][and references
therein]{1994ApJ...429..300W}.

Our analysis of the X-ray data suggests that
\casa\ has been expanding in a dense RSG wind whose swept-up mass
is around 6 \msun.
This large mass of swept-up RSG wind seems to indicate that the mass
loss of the \casa's progenitor might have been primarily via the RSG wind.
However,
stellar evolution models, with commonly
adopted mass-loss rates for RSG stars
\citep[e.g.,][]{1988A&AS...72..259D}, predict significantly less
mass loss via the RSG wind
for a single $\sim$16 \msun\ star \citep[e.g.,][]{RevModPhys.74.1015}.
For
the progenitor of SN~1993J, the mass loss is attributed to Roche lobe
overflow in the binary system \citep[][and references
therein]{Stancliffe09_MNRAS396.p1699}. This was supported by the
possible existence of a binary companion \citep{2004Natur.427..129M,2009Sci...324..486M}.
A similar binary system also has been suggested for \casa\
\citep[e.g.,][]{Young06_640.p891}, although
no companion star has currently been identified.
We note, however, that the mass loss rate of cool stars such as
RSG stars is largely
uncertain, and models may have underestimated
the mass loss during the RSG phase.  For example, stellar pulsations
may trigger an unusually strong wind which substantially increases the mass
loss rate in the RSG phase. \citet{Yoon10_717.L62} suggested that, if
that happens, a single 20 \msun\ star can explode as a Type~IIb SN
after losing most of its hydrogen envelope.

\section{Cosmic Ray Acceleration}
\label{sec:discuss-cr}

So far we have neglected effects of cosmic ray acceleration in our analysis.
However, the existence of X-ray nonthermal filaments around \casa\
clearly indicates that
electrons are accelerated up to the TeV energies.
Observations also suggest that magnetic fields are amplified at the shock,
which suggests the acceleration of
cosmic ray protons \citep[see ][for a review]{2008ARA&A..46...89R}.
Modeling of the broadband nonthermal
emission from the radio to the gamma-ray bands indicates acceleration
of protons as well
\citep[e.g.,][]{2003A&A...400..971B}.
Inclusion of cosmic ray acceleration will change the self-similar
solutions we used, where a principal effect of efficient cosmic ray
acceleration would be a reduction of the distances between the forward
shock and the reverse shock, as demonstrated in the Tycho SNR
by \citet{2005ApJ...634..376W}. However,
the tendency of increasing density and decreasing temperature toward
the contact discontinuity is not likely affected
\citep{2000ApJ...543L..57D}.
Thus, our conclusion that \casa\ is expanding inside a wind remains
valid even though cosmic acceleration is efficient in \casa.
On the other hand,
we might have overestimated the wind density, as the cosmic ray
acceleration can increase the postshock gas compression ratio. For a
strong relativistic shock, the compression ratio would be 7. The
compression ratio will be further increased if accelerated cosmic rays
efficiently escape the system.

We have assumed that the thermal emission we analyzed is from the
postshock gas. In particular, we assumed that those regions we used
for azimuthal analysis represent postshock gas right behind the
forward shock. However, most of these regions are located beyond the
nonthermal filaments that are the locations of forward
shocks. We note that the intensity of the nonthermal filaments varies
along the boundary, and it is particularly faint in the western
boundary where the thermal emission is clearly seen. In this regard,
we consider that the thermal emission beyond the nonthermal
filaments are likely a projection of different shock fronts.
\mynoteB{
It indicates that not all outer shocks of {\casa} accelerate
cosmic ray electrons efficiently enough to show X-ray nonthermal
filaments.
An interesting possibility is that the mean expansion velocity of the
shock front with nonthermal X-ray emission is slower than those
without nonthermal X-ray emission.
Understanding
differences of these shocks may provide valuable insight on the
cosmic ray acceleration in the shocks.
}

\section{Summary}
\label{sec:summary-nature}

Using a 1 Msec \emph{Chandra} observation, we present our detailed analysis of
the blast wave in the Galactic SNR \casa.
By analyzing the X-ray emission from the shocked ambient gas, we have
studied the nature of the ambient medium around \casa, and also the
nature of the progenitor star.  Our results provide observational
evidence of \casa\ interacting with the RSG winds from the progenitor
star.

\begin{itemize}
\item After removing spatially-varying background components, we
  identified faint emission, with no prominent limb-brightening, around
  the outer boundary of \casa.
\item X-ray spectra of these regions are well fit with the emission
  from a thermal plasma of subsolar abundances, indicating
  they are shocked ambient gas.
\item In the western boundary where nonthermal emission is
  relatively weak, we extracted a radial series of spectra.
  Comparison
  with self-similar solutions of \citet{1982ApJ...258..790C} indicates that
  \casa\ is currently expanding inside a wind, rather
  than a uniform medium.
\item The postshock density estimates from the spectra of the outermost
  shocked wind range between $3$ and $5$ \cmthree, implying a preshock
  wind density of $\sim 1$ \cmthree\ at the current outer radius of
  \casa\ ($\sim 3$ pc). The high density of the wind suggests that it
  is a dense wind from a red supergiant (RSG) star.
\item Combined with results of \citet{2003ApJ...593L..23C}, our
  measurements suggest that the ejected mass during the SN explosion
  was about $\sim 4\,\msun$, implying total mass of the star
  (including the mass of the neutron star) at the explosion of $\sim5\
  \msun$. Since \casa\ was Type~IIb SN \citep{2008Sci...320.1195K}, we
  estimate $\sim16\ \msun$ as the initial mass of the progenitor star.
\item Our estimated explosion energy of $5 \times 10^{51}$ erg is larger than
  that of normal core-collapse SNe ($\sim10^{51}$ erg), although not unprecedented
  for Type IIb SNe \citep{2009ApJ...703.1612H}. On the other hand, the explosion
  energy would have been overestimated if wind density is overestimated, and
  the uncertainties in our estimates allow for a lower explosion energy.
\item Our results suggest that the swept-up mass of the RSG wind is
  about 6 \msun. This indicates that the progenitor of
  \casa\ may have lost its mass primarily via the RSG wind.
\end{itemize}

\nolinenumbers

\acknowledgements


\end{document}